\documentclass[12pt]{article}
\usepackage{amsmath,amssymb,amsthm,epsfig,color}
 \usepackage{hyperref}  

 \newcommand{\swnedots}{\mathinner{\mkern1mu\raise1pt\vbox{\kern7pt
       \hbox{.}}\mkern2mu
     \raise4pt\hbox{.}\mkern2mu\raise7pt\hbox{.}\mkern1mu}}

\renewcommand{\(}{\left(}
\renewcommand{\)}{\right)}

\DeclareMathOperator{\Exp}{Exp}
\DeclareMathOperator{\Vect}{Vect}
\DeclareMathOperator{\Dim}{dim} 
\DeclareMathOperator{\Spec}{Spec}
\DeclareMathOperator{\Int}{Int}

\newcommand{\abs}[1]{\left\lvert#1\right\lvert}

\newtheorem{prop}{Proposition}[section]
\newtheorem{theo}{Theorem}
\newtheorem{lemma}[prop]{Lemma}
\newtheorem{remark}[prop]{Remark} 
\newtheorem{ex}[prop]{Example} 
\newtheorem{corol}[prop]{Corollary}
\newtheorem{defi}[prop]{Definition}
 
\numberwithin{equation}{section}

\newcommand{\surj}{\to\kern-.1em\llap{$\to$}}

\newcommand{\jrus}{\leftarrow\kern-.1em\llap{$\leftarrow$}}
\newcommand{\strictsubset}{\hbox{$\subseteq\kern-.4em\llap{${}_/$}$}}

\begin{document}
\title{{Exponentials form a basis of discrete holomorphic functions}}
\author{Christian \textsc{Mercat}\\
\small
 Sonderforschungsbereich 288, MA 8-5\\
\small
Technische Universit\"at Berlin\\
\small
Stra\ss e des 17. Juni 136, D-10623 Berlin, Germany\\
\small
\href{mailto:mercat@sfb288.math.tu-berlin.de}{mercat@sfb288.math.tu-berlin.de}}
\maketitle
\begin{abstract} We show that discrete exponentials form a basis of discrete
  holomorphic functions. On a convex, the discrete polynomials form a basis
  as well.
\end{abstract}

\section{Introduction}
The notion of discrete Riemann surfaces has been defined in~\cite{M01}.  We
are interested in a cellular decomposition $\diamondsuit$ of the complex
plane or a simply connected portion $U$ of it, by \emph{rhombi}. We have a map
from the vertices $\diamondsuit_0$ to the complex plane
$Z:\diamondsuit_0\to\mathbb{C}$ such that for each oriented face
$(x,y,x',y')\in\diamondsuit_2$, its image is a positively oriented rhombus
$\(Z(x),Z(y),Z(x'),Z(y')\)$ of side length $\delta>0$. It defines a
straightforward Cauchy-Riemann equation for a function $f\in
C^0(\diamondsuit)$ of the vertices, and similarly for forms:
\begin{equation}
  \label{eq:CR}
  \frac{f(y')-f(y)}{Z(y')-Z(y)}=\frac{f(x')-f(x)}{Z(x')-Z(x)}.
\end{equation}

\begin{figure}[htbp]
\begin{center}\input{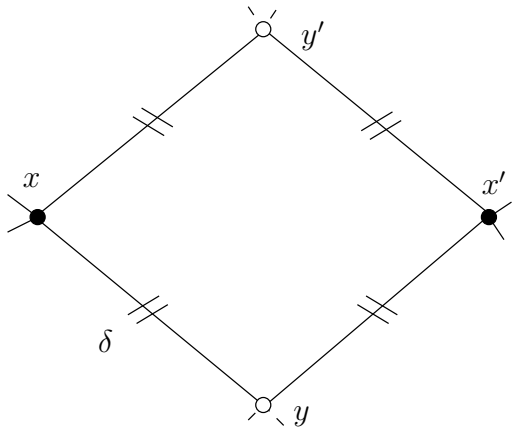}
\end{center}
\caption{The discrete Cauchy-Riemann equation takes place on each rhombus.}         \label{fig:CR}
\end{figure}

We call such a data a \emph{critical map} of $U$. The relevance of this kind
of maps in the context of discrete holomorphy was first pointed out and put
to use by Duffin~\cite{Duf68}. The rhombi can be split in four, yielding a
finer critical map $\diamondsuit',\, Z'$ with $\delta'=\delta/2$.
In~\cite{M01, M0206041} we proved that a converging sequence of discrete
holomorphic functions on a refining sequence of critical maps converges to a
continuous holomorphic function and any holomorphic function on $U$ can be
approximated by a converging sequence of discrete holomorphic functions. The
proof was based on discrete polynomials and series. In the present article we
are going to show that the vector space spanned by discrete polynomials is
the same as the one spanned by discrete exponentials and the main result is
the following
\begin{theo}\label{prop:ExpBasCvx}
  On a combinatorially convex map, the discrete exponentials form a basis of
  the discrete holomorphic functions.
\end{theo}
On a non combinatorially convex map we define some special exponentials which
supplement this basis.

The article is organized as follows. After recalling some basic features of
discrete Riemann surfaces at criticality in Sec.~\ref{sec:IntDer}, we define
discrete exponentials in Sec.~\ref{sec:Exp} and show some of its basic
properties, related to polynomials and series. We give in particular a
formula for the expression of a generic exponential in a basis of
exponentials. In Sec.~\ref{sec:ConvexityTT}, we introduce the notion of
convexity, related to a geometrical construction called train-tracks, and we
prove the main result. Finally we study the general case in
Sec.~\ref{sec:SpecExp} where we define special exponentials and show they
form a basis. The appendix lists some other interesting properties of the
discrete exponentials which are not needed in the proof.

We note that it is possible, in the genus $1$ case, to use the wonderful
machinery defined in~\cite{Ken02} to prove that discrete exponentials
form a basis of discrete holomorphic functions. Indeed, Richard Kenyon gives
an expression of the discrete Green's function as an integral over the space
of discrete exponentials:
\begin{equation}
  \label{eq:KenyonGreen}
  G(O,x)=-\frac{\log \frac{\delta}{2}}{8\,\pi^2\,i}\oint_{C}\,
\Exp({:}\lambda{:}\,x) \, \frac{\log \lambda}{\lambda}\, d\lambda
\end{equation}
where the integration contour $C$ contains all the points in $P_\diamondsuit$
(the possible poles of $\Exp({:}\lambda{:}\,x)$) but avoids the negative real
line.

\section{Integration and Derivation at criticality}\label{sec:IntDer}
\subsection{Integration} \label{sec:Integration}
Given an isometric local map $Z:U\cap\diamondsuit\to\mathbb{C}$, where the image
of the quadrilaterals are lozenges in $\mathbb{C}$, any holomorphic function
$f\in\Omega(\diamondsuit)$ gives rise to an holomorphic $1$-form $f dZ$ defined by
the formula,
\begin{equation}
  \label{eq:fdZ}
  \int_{(x,y)} f dZ :=\frac{f(x)+f(y)}2 \(Z(y)-Z(x)\),
\end{equation}
where $(x,y)\in\diamondsuit_1$ is an edge of a lozenge. It fulfills the
Cauchy-Riemann equation for forms which is, in the same conditions as Eq.~\eqref{eq:CR}:
\begin{eqnarray}
\frac{1}{Z(y')-Z(y)}\,\,
\(
\int\limits_{(y,x)}\!\!+\!\!\int\limits_{(x,y')}
\!+\!\int\limits_{(y,x')}\!\!+\!\!\int\limits_{(x',y')}
\)&&
\,f\,dZ  
  \label{eq:CRforms}
\\
\qquad =
\frac{1}{Z(x')-Z(x)} &&
\(
\int\limits_{(x,y)}\!\!+\!\!\int\limits_{(y,x')}
\!+\!\int\limits_{(x,y')}\!\!+\!\!\int\limits_{(y',x')}
\)
\,f\,dZ.
\notag
\end{eqnarray}

Once an origin $O$ is chosen, it provides a way to integrate a function
$\Int\,(f)(z):=\int_O^zf dZ$.  We proved in~\cite{M0206041} that the
integrals of converging discrete holomorphic functions $(f^k)_k$ on a
refining sequence $(\diamondsuit^k)_k$ of critical maps of a compact converge to the
integral of the limit. If the original limit was of order
$f(z)=f^k(z)+O(\delta_k^2)$, it stays this way for the integrals,
$\int_O^zf(u)\,du=\int_O^zf^k\,dZ+O(\delta_k^2)$, where the left hand side is
the usual continuous integral and the right hand side the discrete ones.

Following Duffin~\cite{Duf, Duf68}, we define by inductive integration the
discrete analogues of the integer power monomials $z^k$, that we denote
$Z^{:k:}$:
\begin{eqnarray}
  \label{eq:Zk}
  Z^{:0:}&:=&1,\\
  Z^{:k:}&:=&k\,\int_{O} \,Z^{:k-1:}\, dZ.
\end{eqnarray}

The discrete polynomials of degree less than three agree point-wise with
their continuous counterpart, $Z^{:2:}(x)=Z(x)^2$ so that by repeated
integration, the discrete polynomials in a refining sequence of a compact
converge to the continuous ones and the limit is of order $O(\delta^2)$.

\subsection{Derivation} \label{sec:Derivation}
The combinatorial surface being simply connected and the graph $\diamondsuit$
having only quadrilateral faces, it is bi-colorable. Let $\Gamma$ and
$\Gamma^*$ the two sets of vertices and $\varepsilon$ be the
\emph{biconstant} $\varepsilon(\Gamma)=+1$, $\varepsilon(\Gamma^*)=-1$.  For
a holomorphic function $f$, the equality $f\,dZ\equiv 0$ is equivalent to
$f=\lambda\,\varepsilon$ for some $\lambda\in \mathbb{C}$.

Following Duffin~\cite{Duf, Duf68}, we introduce the
\begin{defi}\label{def:fp}
  For a holomorphic function $f$, define on a flat simply connected map $U$
  the holomorphic functions $f^\dag$, the \emph{dual} of $f$, and $f'$, the
  \emph{derivative} of $f$, by the following formulae:
  \begin{equation}
f^\dag(z):=\varepsilon(z)\,\bar f(z),\label{eq:fdag}
\end{equation}
where $\bar f$ denotes the
  complex conjugate, $\varepsilon=\pm 1$ is the biconstant, and
  \begin{equation}
f'(z):=\frac4{\delta^2}\left( \int_{O}^z f^\dag
    dZ\right)^\dag+\lambda\,\varepsilon,\label{eq:deffp}
  \end{equation}
defined up to $\varepsilon$.
\end{defi}

We proved  in~\cite{M01} the following 
\begin{prop}\label{prop:fp}
  The derivative $f'$ fulfills
\begin{equation}
  \label{eq:dffpdz}
  d\, f = f'\, dZ.
\end{equation}
\end{prop}

\section{Exponential} \label{sec:Exp}
\subsection{Definition}\label{sec:ExpDef}
\begin{defi}
  The discrete exponential $\Exp({:}\lambda{:}\,Z)$ is the solution of
\begin{eqnarray}
\Exp({:}\lambda{:}\,O)&=&1\label{eq:Exp}\\
d\,\Exp({:}\lambda{:}\,Z)&=&\lambda\,\Exp({:}\lambda{:}\,Z)\,dZ\label{eq:dExp}
\end{eqnarray}
\end{defi}

It was first defined in~\cite{M0111043} and put to a very interesting use
in~\cite{Ken02}.  For $\abs{\lambda}\not=2/\delta$, an immediate check shows that
it is a rational fraction in $\lambda$ at every point,
\begin{equation}
  \label{eq:ExpRatFrac}
  \Exp({:}\lambda{:}\,x)=\prod_{k}
\frac{1+\frac{\lambda\delta}{2}e^{i\,\theta_k}}{
1-\frac{\lambda\delta}{2}e^{i\,\theta_k}}
\end{equation}
where $(\theta_k)$ are the angles defining $(\delta\,e^{i\,\theta_k})$, the
set of ($Z$-images of) $\diamondsuit$-edges between
$x=\sum\delta\,e^{i\,\theta_k}$ and the origin. Because the map is critical,
Eq.~\eqref{eq:ExpRatFrac} only depends on the end points $(O,x)$.

\begin{prop}\label{prop:ExpDag}
For point-wise multiplication, at every point
$x\in\diamondsuit_0$, 
\begin{equation}
\Exp({:}\lambda{:}\,x)\cdot\Exp({:}-\lambda{:}\,x)=1.\label{eq:ExpExp1}
\end{equation}

  The anti-linear duality $\dag$ maps exponentials to exponentials:
\begin{equation}
  \label{eq:ExpDag}
  \Exp({:}\lambda{:})^\dag=\Exp({:}\frac{4}{\delta^2\bar\lambda}{:}).
\end{equation}
\end{prop}
In particular, $ \Exp({:}\infty{:})=1^\dag=\varepsilon$ is the biconstant.
\begin{proof}[Proof \ref{prop:ExpDag}]
The first assertion is immediate.

Derivation of $\Exp({:}\lambda{:})^\dag$ gives, 
\begin{eqnarray}
  \label{eq:ExpDagDeriv}
\(\Exp({:}\lambda{:})^\dag\)'&=&
\frac{4}{\delta^2}\(\int_O^z \Exp({:}\lambda{:})\, dZ\)^\dag+\mu\,\varepsilon\\
&=&\frac{4}{\delta^2}\,\(\frac{\Exp({:}\lambda{:})-1}{\lambda}\)^\dag
+\mu\,\varepsilon
\notag\\
&=&\frac{4}{\delta^2\bar\lambda}\,\Exp({:}\lambda{:})^\dag+\nu\,\varepsilon
\notag
\end{eqnarray}
with $\mu,\,\nu$ some constants, so that the initial condition
$\Exp({:}\lambda{:}O)^\dag=1$ at the origin and the difference equation
$d\,\Exp({:}\lambda{:})^\dag=
\frac{4}{\delta^2\bar\lambda}\,\Exp({:}\lambda{:})^\dag\,dZ$ yields the
result.
\end{proof}

Note that it is natural to define
$\Exp({:}\lambda{:}(x-x_0)):=\frac{\Exp({:}\lambda{:}x)}{\Exp({:}\lambda{:}x_0)}$
as a function of $x$ with $x_0$ a fixed vertex. It is simply a change of
origin. But apart on a lattice where addition of vertices or multiplication
by an integer can be given a meaning as endomorphisms of the lattice, there
is no easy way to generalize this construction to other discrete holomorphic
functions such as $\Exp({:}\lambda{:}(x+n\,y))$ with $x,y\in\diamondsuit_0$
and $n\in\mathbb{Z}$.

\subsection{Series}\label{sec:Series}
The series $\sum_{k=0}^{\infty}\frac{\lambda^{k}\,Z^{:k:}}{k!}$, wherever it
is absolutely convergent, coincide with the rational
fraction~\eqref{eq:ExpRatFrac}: Its value at the origin is $1$ and it
fulfills the defining difference equation~\eqref{eq:dExp}. Because $\frac{Z^{:k:}}{k!}$ are
the iterations of the integration operator $\Int$ on the constant function
$1$, their norm can not grow faster than the powers of its largest eigenvalue
$\lambda_{\text{max}}$, which implies absolute convergence for
$\abs{\lambda}<1/\abs{\lambda_{\text{max}}}$. We have some information on
these eigenvalues, summarized in the appendix, through the minimal polynomial of
$Z$. Direct analysis gives an estimate of $Z^{:k:}$:
\begin{prop}\label{prop:ZkCrois}
  For $x\in\diamondsuit$, at a combinatorial distance $d(x,O)$ of the origin,
  and any $k\in\mathbb{N}$,
  \begin{equation}
    \label{eq:ZkCrois}
    \abs{\frac{Z^{:k:}}{k!}}\leq\(\frac{\alpha+1}{\alpha-1}\)^{{d({x},O)}}
\(\alpha\,\frac{\delta}{2}\)^k,
  \end{equation}
for any $\alpha>1$ arbitrarily close to $1$.
\end{prop}

\begin{corol}\label{prop:ExpSeriesConv}
  The series  $\sum_{k=0}^{\infty}\frac{\lambda^{k}\,Z^{:k:}}{k!}$ is
  absolutely convergent for $\abs{\lambda}<\frac{2}{\delta}$.
\end{corol}

\begin{proof}[Proof \ref{prop:ZkCrois}]
  It is proved by double induction, on the degree $k$ and on the
  combinatorial distance to the origin.  
  
  For $k=0$, it is valid for any $x$ since
  $\frac{\alpha+1}{\alpha-1}=1+\frac{2}{\alpha-1}>1$, with equality only at
  the origin.
  
  Consider $x\in\diamondsuit$ a neighbor of the origin,
  $Z(x)=\delta\,e^{i\,\theta}$, then an immediate induction gives for $k\geq
  1$,
  \begin{equation}
    \label{eq:ZkxO}
    \frac{Z^{:k:}(x)}{k!}=2\(\frac{\delta\,e^{i\,\theta}}{2}\)^k
  \end{equation}
  which fulfills the condition Eq.~\eqref{eq:ZkCrois} for any $k\geq1$
  because $\frac{\alpha+1}{\alpha-1}\,\alpha^k>2$. This was done merely for
  illustration purposes since it is sufficient to check that the condition
  holds at the origin, which it obviously does.  
  
  Suppose the condition is satisfied for a vertex $x$ up to degree $k$, and
  for its neighbor $y$, one edge further from the origin, up to degree $k-1$.
  Then,
\begin{equation}
  \label{eq:Zkxy}
  \frac{Z^{:k:}(y)}{k!}=\frac{Z^{:k:}(x)}{k!}+
\frac{Z^{:k-1:}(x)+Z^{:k-1:}(y)}{(k-1)!}\frac{Z(y)-Z(x)}{2}
\end{equation}
 in absolute value fulfills
\begin{eqnarray}
  \abs{\frac{Z^{:k:}(y)}{k!}}&\leq&
  \(\frac{\alpha+1}{\alpha-1}\)^{{d({x},O)}}\(\alpha\,\frac{\delta}{2}\)^{k-1}
\(
\(\alpha\,\frac{\delta}{2}\)+
\({1+\frac{\alpha+1}{\alpha-1}}\)\frac{\delta}{2}\)
\notag
\\
&& =\(\frac{\alpha+1}{\alpha-1}\)^{{d({x},O)}}\(\alpha\,\frac{\delta}{2}\)^{k}
\(1+\frac{2}{\alpha-1}\)  \label{eq:ZkxyAbs}
\\
&& =\(\frac{\alpha+1}{\alpha-1}\)^{{d({y},O)}}\(\alpha\,\frac{\delta}{2}\)^{k},
\notag
\end{eqnarray}
thus proving the condition for $y$ at degree $k$. It follows by induction
that the condition holds at any point and any degree.
\end{proof}

\subsection{Basis}\label{sec:Basis}
\begin{theo}\label{prop:ExpZk}
  The vector spaces of discrete exponentials and of
  discrete polynomials coincide. A basis is given by any set of exponentials
  $\left\{\Exp({:}\lambda_\ell{:})\right\}_{1\leq \ell\leq n}$ of the right
  dimension $n=\Dim\Vect\(Z^{:k:}\)$, with
  $\lambda_k\not=\lambda_\ell$, distinct complex values of norm different
  from $\frac{2}{\delta}$.
In such a basis, if for a fixed $\lambda_0$, 
  \begin{equation}
\Exp({:}\lambda_0{:})=\sum_{\ell=1}^n
  \mu_\ell(\lambda_0)\,\Exp({:}\lambda_\ell{:}),\label{eq:Exp0SExp}
\end{equation}
then $\Exp({:}\lambda{:})$ is expressed as
\begin{equation}
  \label{eq:ExpSExp}
  \Exp({:}\lambda{:})=\(
\sum_{\ell=1}^n\frac{\lambda_0-\lambda_\ell}{\lambda-\lambda_\ell}\,\mu_\ell(\lambda_0)
\)^{-1}
\sum_{\ell=1}^n
\frac{\lambda_0-\lambda_\ell}{\lambda-\lambda_\ell}\,\mu_\ell(\lambda_0)\,
\Exp({:}\lambda_\ell{:}).
\end{equation}
\end{theo}
\begin{proof}[Proof \ref{prop:ExpZk}]
  An exponential being equal to a series belongs to the space of polynomials.
  We prove the reciprocal inclusion by induction.
  $Z^{{:}0{:}}=1=\Exp({:}0{:})$.  Suppose
  $Z^{{:}\ell{:}}\in\Vect(\Exp({:}\lambda{:})\,/\,\lambda\in\mathbb{C})$ for
  $\ell<k$, then
\begin{equation}
  \label{eq:ExpLim}
  Z^{{:}k{:}}=\lim_{\lambda\to 0}\lambda^{-k}\left(
\Exp({:}\lambda{:}Z)-
\sum_{\ell=0}^{k-1}\frac{\lambda^{\ell}\,Z^{:\ell:}}{\ell!}
\right)
\end{equation}
is in this compact vector space as well.

Let $\left\{\Exp({:}\lambda_\ell{:})\right\}_{0\leq\ell\leq k}$ a set of non
trivial linear dependent exponentials, $\lambda_\ell\not=\lambda_k$. It can
be reduced by iteration of deletions to a single linear combination
dependence
\begin{equation}
\Exp({:}\lambda_0{:})=\sum_{\ell=1}^k
  \mu_\ell(\lambda_0)\,\Exp({:}\lambda_\ell{:}),\label{eq:Exp0SExp2}
\end{equation}
with $\left\{\Exp({:}\lambda_\ell{:})\right\}_{1\leq\ell\leq k}$ a free set
of at least two elements. We are going to show that it then provides a basis.

We show Eq.~\eqref{eq:ExpSExp} first for $\lambda=0$. The integral of
$\Exp({:}\lambda_0{:})$ is equal to
  \begin{equation}
    \label{eq:ExpPrim}
    \int_O^z \Exp({:}\lambda_0{:})=\frac{\Exp({:}\lambda_0{:})-1}{\lambda_0}
=\sum_{\ell=1}^k
  \mu_\ell(\lambda_0)\,\frac{\Exp({:}\lambda_\ell{:})-1}{\lambda_\ell}
  \end{equation}
so that
\begin{equation}
  \label{eq:ExpPrim2}
  \sum_{\ell=1}^k
  \frac{\lambda_0}{\lambda_\ell}\,\mu_\ell(\lambda_0)\,\Exp({:}\lambda_\ell{:})
=\Exp({:}\lambda_0{:})-1+\sum_{\ell=1}^k
  \frac{\lambda_0}{\lambda_\ell}\,\mu_\ell(\lambda_0).
\end{equation}
One gets as well that
$1-\sum_{\ell=1}^k\frac{\lambda_0}{\lambda_\ell}\,\mu_\ell(\lambda_0)\not=0$
because, using the fact that the LHS exponentials form a free set, the
contrary would imply that
$\frac{\lambda_0}{\lambda_\ell}\,\mu_\ell(\lambda_0)=\mu_\ell(\lambda_0)$ for
all $1\leq\ell\leq k$, which is contradictory.  The RHS of
Eq.~\eqref{eq:ExpSExp} for $\lambda=0$ is
\begin{gather}
  \label{eq:ExpSExp0}
  \(
\sum_{\ell=1}^k\frac{\lambda_0-\lambda_\ell}{-\lambda_\ell}\,\mu_\ell(\lambda_0)
\)^{-1}
\sum_{\ell=1}^k
\frac{\lambda_0-\lambda_\ell}{-\lambda_\ell}\,\mu_\ell(\lambda_0)\,
\Exp({:}\lambda_\ell{:})\qquad\qquad\qquad\\
 =\(
1-\sum_{\ell=1}^k\frac{\lambda_0}{\lambda_\ell}\,\mu_\ell(\lambda_0)
\)^{-1}\(-
\sum_{\ell=1}^k
\frac{\lambda_0}{\lambda_\ell}\,\mu_\ell(\lambda_0)\,
\Exp({:}\lambda_\ell{:})+\sum_{\ell=1}^k\mu_\ell(\lambda_0)\,
\Exp({:}\lambda_\ell{:})\)\notag\\
 =\(
1-\sum_{\ell=1}^k\frac{\lambda_0}{\lambda_\ell}\,\mu_\ell(\lambda_0)
\)^{-1}\(-\Exp({:}\lambda_0{:})+1-
\sum_{\ell=1}^k
\frac{\lambda_0}{\lambda_\ell}\,\mu_\ell(\lambda_0)\,
+\Exp({:}\lambda_0{:})\)=1.\notag
\end{gather}
So that the assertion is proved for $\lambda=0$ and the constants belong to
the vector space spanned by our exponentials. The integration operator on
exponentials, $\Int\(
\Exp({:}\lambda_0{:})\)=\frac{\Exp({:}\lambda_0{:})-1}{\lambda_0}$ preserves
this space therefore, successive integration shows that polynomials belong to
this space as well. We conclude that the free set of exponentials is in fact
a basis of polynomials, hence of exponentials. So there exists $k=n$ complex
valued functions $\mu_\ell(\lambda)$ such that, for all
$\lambda\in\mathbb{C}$ of norm different from $2/\delta$,
  \begin{equation}
\Exp({:}\lambda{:})=\sum_{\ell=1}^n
  \mu_\ell(\lambda)\,\Exp({:}\lambda_\ell{:}).\label{eq:ExpLambdaSExp}
\end{equation}
Therefore the previous considerations can be done for a generic $\lambda$.
This shows in particular that $\mu_\ell(\lambda)\not=0$ for all $\ell$ at a
generic permissible point.  Identifying the expansion of the constant $1$ on
the basis, induced by a generic $\lambda$ and by $\lambda_0$, one gets, for
all $1\leq m\leq n$,
\begin{equation}
  \label{eq:ExpLambdaLambda0}
  \(
\sum_{\ell=1}^n\frac{\lambda-\lambda_\ell}{\lambda_\ell}\,\mu_\ell(\lambda)
\)^{-1}
\frac{\lambda-\lambda_m}{\lambda_m}\,\mu_m(\lambda)
=\(
\sum_{\ell=1}^n\frac{\lambda_0-\lambda_\ell}{\lambda_\ell}\,\mu_\ell(\lambda_0)
\)^{-1}
\frac{\lambda_0-\lambda_m}{\lambda_m}\,\mu_m(\lambda_0),
\end{equation}
so that $(\lambda-\lambda_m)\,\mu_m(\lambda)$ for $1\leq m\leq n$ are all
proportional to the same rational fraction. Solving at the origin yields the
result Eq.~\eqref{eq:ExpSExp}.

It  implies that the value $0$ among the distinct
parameters $(\lambda_\ell)_{0\leq \ell\leq n}$ is permissible as well. By
duality Prop~\ref{prop:ExpDag}, the value $\infty$ yields a well defined
limit as well and in particular
\begin{equation}
  \label{eq:EpsSExp}
  \varepsilon= \(
\sum_{\ell=1}^n({\lambda_0-\lambda_\ell})\,\mu_\ell(\lambda_0)
\)^{-1}
\sum_{\ell=1}^n
({\lambda_0-\lambda_\ell})\,\,\mu_\ell(\lambda_0)\,
\Exp({:}\lambda_\ell{:}).
\end{equation}

As a conclusion, any set of $n$ distinct complex parameters
$\left\{\lambda_k\right\}_k$ of modulus different from $\frac{2}{\delta}$
yields a basis of exponentials. By duality Eq.\eqref{eq:ExpDag}, the argument
$\infty$ can be included as well.
\end{proof}

\begin{remark}
  The derivatives (or integrals) of the discrete holomorphic function $\Exp({:}\lambda{:})$
  with respect to its continuous parameter $\lambda$ yield other interesting
  discrete holomorphic functions which belong to the compact vector space
  spanned by  discrete exponentials. The previous theorem can be restated
  using  as a basis $\(\frac{d^k\,}{d\lambda^k}\Exp({:}\lambda{:})\)_{0\leq
  k<n}$  for any $\lambda\not\in P_\diamondsuit$, since the vector space
  spanned by this set is the same as the limit space spanned by  $\(Exp({:}\lambda_k{:})\)_{1\leq k\leq n}$ 
  when  $\lambda_k$ all converge to $\lambda$, staying   all distinct.
\end{remark}

\section{Convexity}\label{sec:ConvexityTT}
\subsection{Train-tracks}\label{sec:TT}
We are interested only in the compact simply connected case.
\begin{defi} Two opposite sides in a rhombus are parallel. 
  We will call a \emph{train-track} a class of oriented edges of
  $\diamondsuit$ induced by pairing the parallel opposite sides of rhombi.
  Let $\Theta$ this set of classes, each element $t\in\Theta$ is labelled by
  an angle $\theta(t)\in\mathbb{R}\pmod {2\pi}$ defined through any
  representative $e\in t$ by $Z(e)=\delta \exp {i\, \theta(t)}$. The poles of
  the rational fraction Eq.~\eqref{eq:ExpRatFrac} are contained in the set
\begin{equation}
P_\diamondsuit=\frac{2}{\delta}\,\exp {-i\,\theta(\Theta)}.\label{eq:PDiam}
\end{equation}
\end{defi}

There is no problem to define the exponential $\Exp({:}\lambda{:})$ for
$\lambda\not\in P_\diamondsuit$ and the previous results extend to them. In
particular, a basis of discrete exponential is given by any set of
exponentials $\left\{\Exp({:}\lambda_\ell{:})\right\}_{1\leq \ell\leq n}$ of
the right dimension with $\lambda_k\not=\lambda_\ell$, distinct complex (or
infinite) values not in $P_\diamondsuit$.

Prop.~\ref{prop:ExpDag} implies that these exponentials are real on one color
and pure imaginary on the other color, $\Exp({:}\lambda{:})\in\,
C^0_{\mathbb{R}}(\Gamma)+i\,\, C^0_{\mathbb{R}}(\Gamma^*)$, which can be
shown by direct inspection as well.

Each color set $\Gamma$ and $\Gamma^*$ can be given a structure of cellular
decomposition of (a subset of) the domain $U$, their edges are the diagonals
of the lozenges. They are Poincar\'e dual to each other. The Dirichlet
theorem allows to solve for a harmonic real function on the graph $\Gamma$
given some real boundary values. When the weights defining the discrete Laplacian
are given by the aspect ratios of the lozenges (see
Eq.~\eqref{eq:fixApproxHarmoFace} and \cite{M01}), each harmonic function on
one graph can be paired with a dual harmonic function on its dual into a
discrete holomorphic function, uniquely up to an additive constant. Therefore
the dimension of the space of holomorphic functions is equal to half the
number of points on the boundary, plus one~\cite{M01}. And the number of
points on the boundary is also the number of train-tracks, which makes the
set of unoriented train-tracks (and another index) a likely label set for a
basis of discrete holomorphic functions. The construction of this basis is
the subject of Section.~\ref{sec:SpecExp}.

\subsection{Convexity}\label{sec:Convx}
Because the rhombi are all positively oriented, a train-track can not
backtrack, does not self-intersect and two train-tracks can intersect at most
once, or are opposite. A train-track $t\in\Theta$ disconnects the rest of the
cellular decomposition into an initial side and a terminal side. Their role is
exchanged for the opposite train-track $-t$. These sides have two parallel
boundaries facing each other, say $C_i(t)$ and $C_f(t)$, such that
$Z(C_f(t))=\delta\,\exp (i\,\theta(\Theta))+Z(C_i(t))$. A train-track
$t\in\Theta$ identify pair-wise opposite  edges on the positively oriented boundary,
$e^+_t,e^-_t\in\partial \diamondsuit$ with $e^+_t,-e^-_t\in t$. If two
train-tracks $t,t'\in\Theta$ share the same direction,
$\theta(t)=\theta(t')$, the  edges
$e^+_t,e^-_t,e^+_{t'},e^-_{t'}\in\partial \diamondsuit$, with
$Z(e^+_t)=-Z(e^-_t)=Z(e^+_{t'})=-Z(e^-_{t'})$, can occur cyclically in
essentially two ways: Since the train-tracks can not cross, these two pairs are not
interlaced but in the order
\begin{align}
  (e^+_t,\ldots,e^-_t,\ldots,e^+_{t'},\ldots,e^-_{t'},\ldots)&&\text{or}
\label{eq:concav}\\
  (e^+_t,\ldots,e^-_t,\ldots,e^-_{t'},\ldots,e^+_{t'},\ldots)&&
\label{eq:convex}
\end{align}
\begin{figure}[htbp]
\begin{center}\input{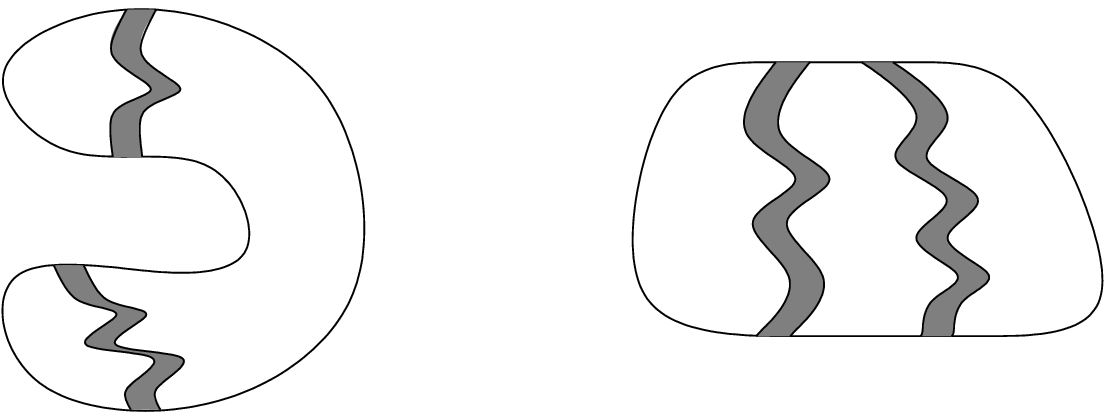}
\end{center}
\caption{Concave and convex situations.}         \label{fig:concavex}
\end{figure}

\begin{defi}\label{def:convex}
  A map $\diamondsuit$ is  \emph{convex} iff  the situation
  \eqref{eq:convex} only occurs.
\end{defi}

\begin{figure}[htbp]
\begin{center}\input{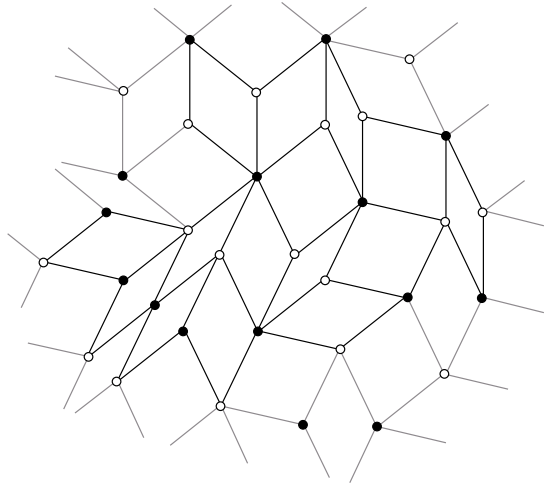}
\end{center}
\caption{An example of a convex.}         \label{fig:convex}
\end{figure}

Be careful that this notion of convexity, while connected to the usual one is
nevertheless different, in particular if no two train-tracks share the same
slope then it is always convex, see for example Fig.~\ref{fig:convex}. This
notion is closely related to the spirit of~\cite{CdV96}.

\begin{prop}\label{prop:convex}
  When $\diamondsuit\subset \diamondsuit'$ is a portion of a larger convex
  compact flat critical map, convexity is equivalent to the fact that if a
  train-track has two rhombi in $\diamondsuit$, then the whole portion in
  between is in $\diamondsuit$ as well.
\end{prop}
\begin{proof}[Proof \ref{prop:convex}]
  Necessity is clear, if a train-track has two disconnected portions in
  the connected $\diamondsuit$, then the latter is not convex. Conversely,
  if every train-track intersects $\diamondsuit$ in a connected line, the
  fact that train-tracks with equal angles don't intersect allows to continue
  them until the boundary of $\diamondsuit'$, without changing the cyclic order.
\end{proof}

\begin{figure}[htbp]
\begin{center}\input{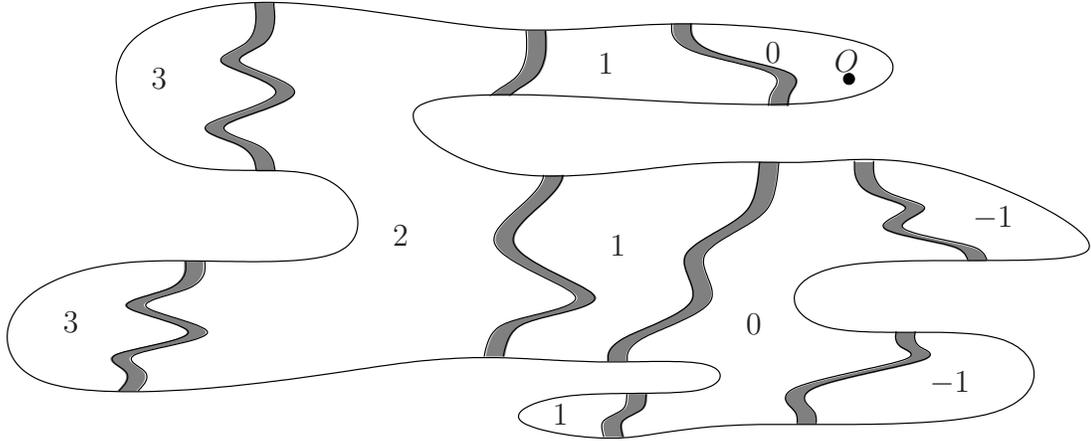}
\end{center}
\caption{The level associated with an angle.}        
\label{fig:TTdeg}
\end{figure}
For future use, we setup the following
\begin{defi}
  Let $t\in\Theta$ a train-track, $\phi=\theta(t)$ its slope angle,
  $\lambda(t)=\frac{2}{\delta}e^{-i\,\phi}$ the corresponding element in
  $P_\diamondsuit$. Let $\diamondsuit^a_\phi, \diamondsuit^b_\phi,\ldots,
  \diamondsuit^z_\phi$ the connected components of $\diamondsuit$ minus the
  rhombi in $\theta^{-1}(\phi)$, having the same slope as $t$. These
  components are ordered into a chain of initial and terminal sides, forming an
  oriented tree\footnote{which is not a total ordering if $\diamondsuit$ is
    not convex.}. Each component has a given level $d_\phi$ counting the
  algebraic number of times these train-tracks were crossed from the
  origin.
To each train-track $t$, we associate the connected component on its terminal
side and denote it  $\diamondsuit^t$. We define likewise $d_\phi(t)$.
%
\end{defi}

\begin{remark}
  The rational fraction  $\Exp({:}\mu{:}\,x)$, for  $x\in\diamondsuit^t$ has a  pole of  order $d_\phi(t)$  at
  $\mu=\lambda(t)$ and a zero of same order at  $\mu=-\lambda(t)$ if $d_\phi(t)>0$, or a zero  of  order
  $-d_\phi$ at $\mu=\lambda(t)$ and a pole of same order at  $\mu=-\lambda(t)$ if $d_\phi(t)<0$.
  
  The level associated with an angle $\phi$ is opposite to the level
  associated with $-\phi$, $d_{-\phi}=-d_\phi$: A zero at $\mu=\lambda_0$
  for $\Exp({:}\mu{:}\,x)$ is equivalent to a pole of the same order at
  $\mu=-\lambda_0$. The two connected components which are separated by a
  given train-track $t$ are $\diamondsuit^t$ and
  $\diamondsuit^{-t}$. See Fig.~\ref{fig:TTdeg} for an example.
\end{remark}

\subsection{Proof of the main result}\label{sec:proof}
\begin{proof}[Proof \ref{prop:ExpBasCvx}]
  Consider a point $x\in\diamondsuit_0$ and a path $\gamma$ form the origin
  to $x$. This path can be so chosen that the train-tracks it crosses are
  crossed only once. Its length is the combinatorial distance $d(x,O)$
  between $x$ and the origin. The point  $x$ is uniquely determined by the set $\Theta_x$ of
  train-tracks the path crosses. Indeed,
  \begin{equation}
    \label{eq:xTT}
    x=\sum_{t\in \Theta_x} \delta\,e^{i\,\theta(t)}
  \end{equation}
  where we have written $x$ in place of its $Z$-image $Z(x)$. Given a basis
  $\left\{\Exp({:}\lambda_\ell{:})\right\}_{1\leq \ell\leq n}$ of discrete  exponentials such as in Prop.~\ref{prop:ExpZk}, 
  the rational fraction  Eq.~\eqref{eq:ExpRatFrac} in $\lambda$, equals the basis expansion
  Eq.~\eqref{eq:ExpSExp},
\begin{eqnarray}
  \label{eq:RFExpSExp}
   \Exp({:}\lambda{:}\,x)&=&\prod_{t\in\Theta_x}
\frac{1+\frac{\lambda\delta}{2}e^{i\,\theta(t)}}{
1-\frac{\lambda\delta}{2}e^{i\,\theta(t)}}\\
&=&\(
\sum_{\ell=1}^n\frac{\lambda_0-\lambda_\ell}{\lambda-\lambda_\ell}\,
\mu_\ell(\lambda_0)
\)^{-1}
\sum_{\ell=1}^n
\frac{\lambda_0-\lambda_\ell}{\lambda-\lambda_\ell}\,\mu_\ell(\lambda_0)\,
\Exp({:}\lambda_\ell{:}\,x).\notag
\end{eqnarray}
The dependency in $x$ of the basis expansion is solely through the complex
numbers $\Exp({:}\lambda_\ell{:}\,x)$. It is immediate to see that the poles
in front of them are canceled by zeros of the prefactor so that 
\begin{equation}
  \label{eq:ExpRatFracx}
  \Exp({:}\lambda{:}\,x)=\frac{P_x(\lambda)}{Q(\lambda)}
\end{equation}
with $P_x$ and $Q$ are polynomials, both of degree $n-1$, the former depending on $x$ but not
the latter. Some zeros of $P_x$ cancel some of the zeros of $Q$, leaving
$d(x,O)$ poles, counted with multiplicities. Each pole is of the form
$\lambda=\frac{2}{\delta}e^{-i\,\theta(t)}$ and corresponds to one of the train-tracks $t$ between
$x$ and the origin.

If the train-tracks angles are all different, considering all the paths
starting at the origin, we see that each train-track such that the origin is
on its initial side contributes to a zero of $Q$. Therefore its degree is at
least the number of such train-tracks. This number is equal to the unoriented
train-tracks, that is to say half the number of vertices on the boundary.
Therefore
\begin{equation}
  \label{eq:dimBasis}
  n\geq\frac{\abs{\partial \diamondsuit}}{2}+1.
\end{equation}
The right hand side is the dimension of the space of discrete holomorphic
functions, which is an upper bound for $n$, yielding the result in that case.

If some train-tracks share the same angle, they don't contribute to different
zeros of $Q$ but possibly to a higher order for the same zero. 

It is a matter of definition to check that when a point $x$ is in a connected
component $\diamondsuit^\ell_\phi$ at a level $d_\phi> 0$, the train-tracks
of angle $\phi$ encountered from the origin, contribute to a zero at
$\lambda=\frac{2}{\delta}e^{-i\,\phi}$ of order $d_\phi$ in $Q(\lambda)$.  When
$d_\phi<0$, the zero is at $\lambda=-\frac{2}{\delta}e^{-i\,\phi}$.
  
  Convexity is equivalent to the fact that there is, for each possible slope,
  only one connected component at a given level. Therefore, in that case as
  well, each (unoriented) train-track contributes to exactly one zero of the
  numerator $Q$, counting multiplicities. The same counting yields the
  result.
\end{proof}

\section{Special Exponentials}\label{sec:SpecExp}
We are going to define the \emph{special exponentials}
$\Exp_t({:}\lambda_0{:}\,x)$ for $\lambda_0\in
P_\diamondsuit$, one for every train-track $t$ of slope $2/\lambda_0=\delta\,e^{i\,\theta(t)}$.

\subsection{Definition}
Let $t\in\Theta$ a train-track, $\phi=\theta(t)$ its slope angle, $\lambda_0=\lambda(t)\in\, P_\diamondsuit$, and
$\diamondsuit^a_\phi, \diamondsuit^b_\phi,\ldots, \diamondsuit^z_\phi$ the associated connected components of $\diamondsuit$.

Consider the rational fraction $\Exp({:}\lambda{:}\,x)$, at a point
$x\in\diamondsuit^\ell_\phi$ on a connected component at level $d_\phi(\ell)>0$.  Choose an origin $O^\ell_\phi\in\diamondsuit^\ell_\phi$
for each connected component $\ell$.  It defines $\Exp_\ell({:}\lambda{:}\,x):=\Exp({:}\lambda{:}\,(x-O^\ell_\phi))|_{\diamondsuit^\ell_\phi}$,
free of zeros and poles at $\lambda=\pm\lambda_0$ for every point $x\in\diamondsuit^\ell_\phi$. We extend it by 
zero to other connected components. The exponential $\Exp({:}\lambda{:}\,(x-O^\ell_\phi))$ has a well
defined limit on the components of level lower than $d_\phi(\ell)$, it is the
null function because on each component it contains a factor of the form
$(\lambda-\lambda_0)$, therefore the Cauchy-Riemann equation is fulfilled
across the train-track to the previous components. On the next components
however, $\Exp({:}\lambda{:}\,(x-O^\ell_\phi))$ diverges since it contains the factor $(\lambda-\lambda_0)^{-1}$. 
Continuing it by zero yields $\Exp_\ell({:}\lambda{:}\,x)$ which is no longer a discrete holomorphic
function because it fails to fulfill the Cauchy-Riemann equation across the
train-track to the next components.

\begin{prop}\label{prop:ExpSmuExp}
  For every connected component  $\diamondsuit^\ell_\phi$,  there exists a unique
  set of complex numbers $\(\mu_m\)$, one for each connected component $\diamondsuit^m_\phi$
  such that the following piece-wise discrete holomorphic function
  \begin{equation}
    \label{eq:SpecExp}
    \sum_{m}\mu_m\,
\frac{d^{d_\phi(m)-d_\phi(\ell)}\,}{d\lambda^{d_\phi(m)-d_\phi(\ell)}}\,
\Exp_m({:}\lambda{:}\,x)
  \end{equation}
is a well defined discrete holomorphic function for $\lambda=\lambda_0$, on
the whole map $\diamondsuit$, with
\begin{enumerate}
\item $\mu_\ell=1$ and
\item $\mu_m=0$ for every component which can not be reached from
  $\diamondsuit^\ell_\phi$ by a path staying above level $d_\phi(\ell)$.
\end{enumerate}
\end{prop}

\begin{defi}
  We call the previous discrete holomorphic function a \emph{special
  exponential} and denote it $\Exp_t({:}\lambda_0{:}\,x)$ for the train-track $t$ such that
$\diamondsuit^\ell_\phi=\diamondsuit^t$.
\end{defi}

\begin{ex}\label{ex:SpecExp}
  The special exponential corresponding to the central zone at
  level $1$ in Fig.~\ref{fig:TTdeg} will be null on the components at levels
  $0$ and $-1$ but also on the components at level $1$ corresponding to the
  lower leg of the figure. It is equal to the special exponential
  corresponding to the other component at level $1$ on the upper leg, up to a
  global multiplicative factor (and provided the origins in the various
  components are kept the same).
  
  The lattices have few different slopes so that the connected components
  appearing above are reduced to thin areas, see Fig.~\ref{fig:SpecExpTri}
  for an explicit example on the triangular/hexagonal lattice. The extreme
  case is the standard square lattice where they are reduced to lines.
\begin{figure}[htbp]
\begin{center}\input{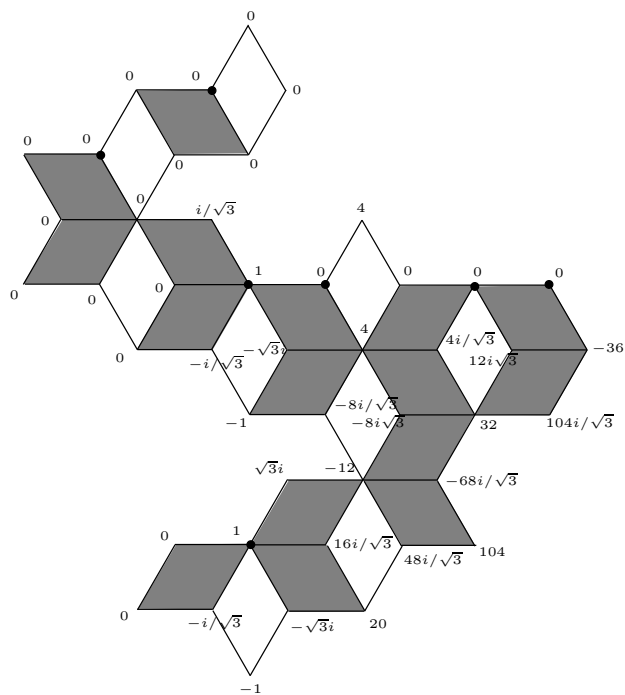}
\end{center}
\caption{A special exponential $\Exp({:}1{:})$ on the standard
  triangular/hexagonal lattice with $\delta=2$. The base point for each
  component is marked by a solid dot and the value at these points is $0$.}
\label{fig:SpecExpTri}
\end{figure}
\end{ex}

The demonstration uses the following
\begin{lemma}\label{prop:ExpSnuExp}
For every connected component $\diamondsuit^\ell_\phi$, and degree $d< d_\phi(\ell)$, there exists a 
unique  set of complex numbers $\(\nu_k\)_{0\leq k< d_\phi(\ell)-d}$  such that
  \begin{equation}
(\lambda-\lambda_0)^d\,\Exp({:}\lambda{:}\,x)-
\sum_{k=0}^{d_\phi(\ell)-d-1}\, \nu_k\, 
\({\lambda-\lambda_0}\)^{d-d_\phi(\ell)+k}
\frac{d^k\,}{d\lambda^k}
\Exp_\ell({:}\lambda{:}\,x)\label{eq:ExpSnuExpl}
\end{equation}
has no poles and no zero at $\lambda=\frac{2}{\delta}e^{-i\,\phi}$ on $\diamondsuit^\ell_\phi$. Moreover, it is proportional to 
$\frac{d^{d_\phi(\ell)-d}\,}{d\lambda^{d_\phi(\ell)-d}}\Exp_\ell({:}\lambda{:}\,x)$.
\end{lemma}

\begin{proof}[Proof \ref{prop:ExpSnuExp}]
Given $x\in \diamondsuit^\ell_\phi$, the rational fraction $(\lambda-\lambda_0)^d\,\Exp({:}\lambda{:}\,x)$ has a
  pole at $\lambda=\frac{2}{\delta}e^{-i\,\phi}$ of order $d_\phi(\ell)-d$
  while $\Exp_\ell({:}\lambda{:}\,x)$ is free of zero or pole there. A Taylor
  expansion at $\lambda=\lambda_0$ of the following product  gives
  \begin{gather}
(\lambda-\lambda_0)^{d_\phi(\ell)}\,\Exp({:}\lambda{:}\,x)=
(\lambda-\lambda_0)^{d_\phi(\ell)}\,\Exp({:}\lambda{:}\,O^\ell_\phi)\,
\Exp_\ell({:}\lambda{:}\,x)
\label{eq:ExpExplTaylor}
\\
= \sum_{k=0}^{d_\phi(\ell)-d} \({\lambda-\lambda_0}\)^{k}\,
F_k(\lambda)\,
\frac{d^k\,}{d\lambda^k}\Exp_\ell({:}\lambda{:}\,x)
+o \({\lambda-\lambda_0}\)^{d_\phi(\ell)-d}\notag
\end{gather}
where $F_k(\lambda)$ is a rational fraction with no pole and no zero at $\lambda=\lambda_0$. Defining 
$\nu_k:=F_k(\lambda_0)$, which does not depend on $x$, one gets the result.
\end{proof}

\begin{proof}[Proof \ref{prop:ExpSmuExp}]
  We use the lemma on every connected component with $d=d_\phi(\ell)$.
  Writing as previously the exponential in sums of different rational
  fractions with poles at $\lambda=\lambda_0$ of various orders, shows that
  each order independently belongs to the space of usual exponentials and
  has to fulfill the Cauchy-Riemann equation when  $\lambda$ tends to $\lambda_0$. 
It is a matter of course inside each
  connected component, but it is also the case across the gaps made by the
  train-tracks between these components. In particular the degree $0$
  provides a well defined discrete holomorphic function, which satisfies the
  condition $1)$. Moreover the exponential  $\Exp\({:}\lambda{:}(x-O^\ell_\phi)\)$ for $x\in\diamondsuit^{m}$ with
  $d_\phi(m)<d_\phi(\ell)$, since it contains the factor  $(\lambda-\lambda_0)$, has zero as a limit when
  $\lambda$ tends to $\lambda_0$ for $x\in\diamondsuit^{m}$ with  $d_\phi(m)<d_\phi(\ell)$. Therefore we can discard 
all the components at level less than $d_\phi(\ell)$ (which correspond to
  integration with respect to $\lambda$). Then every connected subtree of the
  partial ordering of the connected components staying above level
  $d_\phi(\ell)$ can be continued by zero on the rest of the map. Which
  proves the fact that the second requirement can be met.
\end{proof}

A by-product of the proof is the following
\begin{corol}
  For each attained level $d\in d_\phi(\Theta)$, there exists a linear
  combination
\begin{equation}
  \label{eq:ExpSpiExp}
  \sum_{t/\,d_\phi(t)=d}\, \pi_t \,\Exp_t({:}\lambda{:}\,x),
\end{equation}
unique up to a multiplicative constant, which belongs to the space of usual
discrete exponentials.
\end{corol}

\subsection{Basis}
We have now to show that they form a generating set for holomorphic
functions. 

As we pointed out in  the example~\ref{ex:SpecExp}, the special exponentials
associated with different train-tracks may coincide. In order to get the
right dimension, we have to select half of them. We defined
$\diamondsuit^{t}$ as the connected component on the \emph{terminal} side of $t$.
\begin{defi}
  A train-track $t$ is \emph{positively oriented} iff a path of shortest
  length from the origin to a point $x\in\diamondsuit^{t}$ 
  contains an edge in $t$.
\end{defi}
It is equivalent to the fact that a path of shortest length from the origin
to a point $x\in\diamondsuit^{-t}$, the previous component in the direction
of $t$, contains no edge in $t$.

The great interest of such a notion is that when a train-track $t$ is
positively oriented, the associated special exponential
$\Exp_t({:}\lambda{:}\,x)$ is null on the subtree on the initial side of $t$,
which contains the origin. This allows for the construction of a very natural
basis starting from the origin.

\begin{theo}\label{prop:SpecExpBasis}
  The set of all special exponentials for positively oriented train-tracks
  and the constant function $1$ form a basis of discrete holomorphic
  functions.
\end{theo}
  This is done simply by expressing the generic exponential on this basis,
  reconstructing its needed poles out of the special exponentials:

\begin{lemma}\label{prop:ExpSpecExp}
    Let $\Theta^+_O$ the set of positively oriented train-tracks from the
    origin $O$. There exists a unique set of complex numbers $(\kappa_t)_{t\in\Theta^+_O}$ such that
    \begin{equation}
      \label{eq:ExpSpecExp}
      \Exp({:}\mu{:})=1+\sum_{t\in\Theta^+_O}\,\kappa_t\,
\(\mu-\lambda(t)\)^{-d_{\theta(t)}}
\Exp_t({:}\lambda(t){:}).
    \end{equation}
\end{lemma}
\begin{proof}[Proof \ref{prop:ExpSpecExp}]
  The two sides of  Eq.~\eqref{eq:ExpSpecExp} are both discrete holomorphic
  functions for generic $\mu$, and at each point are rational fractions of
  $\mu$ with the same set of poles,  $\{\lambda(t_k)\}$  associated with train-tracks $t_k$ 
  between the origin and $x$, and orders $d_{\theta(t_k)}$, uniquely defining $x=\sum_{t_k}\delta\,
  e^{i\,\theta(t_k)}$. 
  Their value at the origin  are both equal to $1$.  
  
  Consider $\Theta^k_O$ the set of positively oriented train-tracks which can
  be reached in $k$ steps from the origin. Let $\diamondsuit^k$ the
  sub-complex of $\diamondsuit$ spanned by continuous paths from the origin
  whose edges are in $\Theta^k_O$ (we will say \emph{spanned} by
  $\Theta^k_O$) . It is a priori a strict sup-set of the set of vertices at a
  combinatorial distance less than $k+1$ from the origin.
  
  Suppose that we have proven Eq.~\eqref{eq:ExpSpecExp} for the sub-complex
  $\diamondsuit^{d}$, assigning a value $\kappa_t$ to every  $t\in\diamondsuit^{d}$. All the other positively oriented 
  train-tracks are  such that their special exponential is null on 
  $\diamondsuit^k$ therefore it makes sense to restrict  Eq.~\eqref{eq:ExpSpecExp} to $\diamondsuit^k$. 
  The equality obviously holds  for $k=0$ where $\diamondsuit^0=\{O\}$.

  A point  $x\in\diamondsuit^{d+1}\setminus\diamondsuit^{d}$ is characterized by a new  train-track  $t\in\Theta^{d+1}_O\setminus\Theta^{d}_O$. 
  The equation  Eq.~\eqref{eq:ExpSpecExp} provides a linear determination for $\kappa_t$:
  \begin{equation}
    \label{eq:ExpSpecKappa}
    \kappa_t:=\lim_{\mu\to\lambda(t)}  
 \(\mu-\lambda(t)\)^{d_{\theta(t)}}
\(
      \Exp({:}\mu{:})-1-\sum_{u\in\Theta^k_O}\,\kappa_u\,
\(\mu-\lambda(u)\)^{-d_{\theta(u)}}
\Exp_u({:}\lambda(u){:})
\).
  \end{equation}
  Since $\Exp_t({:}\lambda(t){:})$ is null on  $\diamondsuit^k$, Eq.~\eqref{eq:ExpSpecExp} restricted to 
  $\diamondsuit^k$ is not perturbed  by this assignation. The right hand side
  of  Eq.~\eqref{eq:ExpSpecExp} is discrete holomorphic on the set spanned by
  $\Theta^{d}_O\cup\{t\}$. On the other hand, the Cauchy-Riemann equation
  allows to solve uniquely for the value at every other vertex spanned by
  $\Theta^{d}_O\cup\{t\}$, therefore Eq.~\eqref{eq:ExpSpecExp} holds for all  these vertices, 
  showing that the  choice of the vertex $x$ is irrelevant. Another vertex
  $y\in\diamondsuit^{d+1}\setminus\diamondsuit^{d}$ defining another train-track  $t'$ defines $\kappa_{t'}$ in a similar way. 
  If $t\cap t'\in\diamondsuit^{d+1}$, one checks easily that the order with  which $\kappa_t$ and $\kappa_{t'}$ are 
  chosen is irrelevant. It is even more so if their intersection is empty 
  or if this  intersection does not belong to $\diamondsuit^{d+1}$. 
  
  Proceeding for every train-track in  $\diamondsuit^{d+1}\setminus\diamondsuit^{d}$, we assign uniquely a value
  $\kappa_u$ to every train-track $u\in\Theta^{d+1}_O\setminus\Theta^{d}_O$ and prove Eq.~\eqref{eq:ExpSpecExp} 
  for the sub-complex $\diamondsuit^{d+1}$. By induction we prove it for the whole map $\diamondsuit$.
\end{proof}
\begin{proof}[Proof \ref{prop:SpecExpBasis}]
 Uniqueness in the lemma shows that the set of special exponential is a free
 set. The number of positively oriented train-tracks plus one (for the constant
 function $1$) is equal to the dimension of discrete holomorphic functions.
\end{proof}

\appendix

\section*{Additional properties}\label{sec:Prop}
\subsection{Eigenvalues of Integration}\label{sec:Eig}
The polynomials on $\diamondsuit$ are finite dimensional, hence there exists
a minimal degree for which $Z^{:n:}$ is linked with the previous monomials.

\begin{prop}\label{prop:IntSpec}
  Let $P_Z=\sum_{k=1}^n a_k Z^{:k:}=0$ the minimal polynomial of the map
  $\diamondsuit$. The eigenvalues of the integration operator are the roots
  of the polynomial $Q=\sum_{k=1}^n k!\, a_k\, \lambda^{k}$,
  \begin{equation}
    \label{eq:IntSpec}
    \Spec(\Int)=Q^{-1}(0).
  \end{equation}
\end{prop}
\begin{proof}[Proof \ref{prop:IntSpec}]
  In the basis $\(Z^{:k:}/{k!}\)_{0\leq k<n}$, where
  $\Int(Z^{:k:}/{k!})=Z^{:k+1:}/{(k+1)!}$, up to
  $Z^{:n:}/{n!}=-\sum_{k=1}^{n-1}k!\, a_k/n!\,a_n \frac{Z^{:k:}}{k!}$, the
  integration operator has the following matrix representation:
  \begin{equation}
    \label{eq:IntMat}
    \Int=
    \begin{pmatrix}
      0&\hdotsfor{3}&0&0\\
1&\ddots&&&\vdots&-a_1/n!\,a_n\\
0&\,\ddots\,&\,\ddots\,&&\vdots&-2!\, a_2/n!\,a_n\\
\vdots&\ddots&\ddots&\ddots&\vdots&\vdots\\
\vdots&&\ddots&\ddots&0&-(n-2)!\,a_{n-2}/n!\,a_n\\
0&\hdotsfor{2}&0&1&-(n-1)!\,a_{n-1}/n!\,a_n
    \end{pmatrix}
  \end{equation}
  and its characteristic polynomial is $Q$. The minimal polynomial
  $P_Z$ can be  normalized so that $P_Z'(Z)=\varepsilon$, that is to say $a_1=1$. The
  eigenvector associated with $0$ is $\varepsilon$. The eigenvector
  associated with $\lambda\in Q^{-1}(0)$ non null is
  \begin{equation}
    \label{eq:IntVep}
    \sum_{k=0}^{n-1}\(\sum_{\ell=k+1}^n \ell!\,a_\ell\,\lambda^{\ell-k-1}\)
\frac{Z^{:k:}}{k!}.
  \end{equation}
\end{proof}

\subsection{Derivation}\label{sec:ExpDeriv}
Any linear normalization of the degree of freedom $\lambda$ in
Eq.~\eqref{eq:deffp} defines a derivation operator
$\frac{d\phantom{\,Z}}{d\,Z}$. For example 
\begin{equation}
  \label{eq:fixApproxHarmoFace}
  \sum_k \rho(O,x_k)\(\frac{f(y_{k+1})-f(y_k)}{Z(y_{k+1})-Z(y_k)}-\lambda\)=0,
\end{equation}
summed over the quadrilaterals $(O,y_k,x_k,y_{k+1})\in\diamondsuit_2$
adjacent to the origin where
$\rho(O,x_k)=-i\,\frac{Z(y_{k+1})-Z(y_k)}{Z(x_k)-Z(O)}>0$ control their
aspect ratio and are the weights of the discrete Laplacian.  This
normalization states that the derivative at the origin is the mean value of
the nearby face derivatives.

Because the derivation operator can have
only a finite number of different eigenvalues, the derivatives of
exponentials are not always exponentials, so that
\begin{equation}
\Exp({:}\lambda{:}\,Z)\,'=\lambda\,\Exp({:}\lambda{:}\,Z)-f_Z(\lambda)\,\varepsilon
\label{eq:Expp}
\end{equation}
with a non trivial function $f_Z$, depending on the normalization.

\begin{prop}\label{prop:ExpDerivEps}
   Let $(\lambda_1,\ldots,\lambda_n)\in\mathbb{C}^n$
  distinct complex values of norm different from $\frac{2}{\delta}$, which
  define a basis $\left\{\Exp({:}\lambda_k{:})\right\}$ of exponentials on
  $\diamondsuit$ and a certain normalization of the derivation operator, that
  is to say $n$ fixed values $f_Z(\lambda_\ell)\in\mathbb{C}$. Then,
  the normalization function is the following rational fraction:
  \begin{equation}
    \label{eq:ExpDerivEps}
    f_Z(\lambda)=
\sum_{\ell=1}^n
\mu_\ell(\lambda)
\(\lambda-\lambda_\ell+{f_Z(\lambda_\ell)}\)
  \end{equation}
in terms of the coordinates of~  
$\Exp({:}\lambda{:})=\sum_{\ell=1}^n\mu_\ell(\lambda)\Exp({:}\lambda_\ell{:})$
on the basis.
\end{prop}
Each coordinate $\mu_\ell(\lambda)$ is a rational fraction given in
Eq.~\eqref{eq:ExpSExp}.  By Eq.~\eqref{eq:Expp}, the zeros of the rational
fraction $f_Z$, if their norm is different from $2/\delta$, give all the
eigenvectors of the derivation operator.  Uniqueness of the exponential
implies that these eigenvalues are always non degenerate.

In~\cite{M0206041}, we defined a normalization adapted to polynomials
of finite degree. It is the least interesting in that respect since its only
eigenvalue is $0$, the unique eigenvector associated being the constant
$1$. We define a normalization adapted to a basis of exponentials:

\begin{defi} Let $(\lambda_1,\ldots,\lambda_n)\in\mathbb{C}^n$
  distinct complex values of norm different from $\frac{2}{\delta}$, which
  define a basis $\left\{\Exp({:}\lambda_k{:})\right\}$ of exponentials on
  $U$.  We normalize the derivation operator Eq.~\eqref{eq:deffp} such that
  it is diagonal in this basis and null on the orthogonal supplement.
\end{defi}

Because of Eq.~\eqref{eq:ExpLim}, the normalization adapted to polynomials is
associated with a sequence $(\lambda^k_\ell)_{k\in\mathbb{N}}$ of $n$
parameters $\lambda_1^k\gg \lambda_2^k\gg \ldots\gg \lambda_n^k$ converging
to zero at different rates with $k$, for example
$\lambda^k_{\ell}=k^{-\ell}$.

\subsection{Refinement}\label{sec:Refin}
In a refining sequence of critical maps, the exponential behavior with
respect to the continuous variable $\lambda$ is recovered in $O(\delta^2)$:
\begin{equation}
  \label{eq:dExpLambdaO}
  \frac{d\,\Exp({:}\lambda{:}\,x)}{d\,\lambda}
= \Exp({:}\lambda{:}\,x)\,\sum_{k}
\frac{\delta\, e^{i\,\theta_k}}{
1-\left(\frac{\lambda\,\delta}{2}e^{i\,\theta_k}\right)^2}
=x\,\Exp({:}\lambda{:}\,x)+O(\delta^2)
\end{equation}
keeping $\lambda$ and $x$ fixed while $\delta\to 0$. 

In general, the $O(\delta^2)$ character of the convergence for polynomials is
lost for series because the higher the degree, the slower the convergence.
Nevertheless, for exponentials, directly expanding
$\exp(\log(\Exp({:}\lambda{:}\,x)))$ in the same conditions as before, we get
\begin{equation}
\Exp({:}\lambda{:}\,x)=\exp(\lambda\, x)+O(\delta^2).\label{eq:ExpOdelta2}
\end{equation}

\subsection{Change of base point}\label{sec:BasePoint}

The change of base point for a polynomial is not as simple as the Pascal
binomial formula of the continuous case~\cite{M0206041}. Nevertheless, for
exponentials, if $\zeta=a(Z-b)$:
\begin{equation}
    \sum_{k=0}^{\infty}\frac{\lambda^{k}}{k!}\zeta^{:k:}(x)
=\Exp_Z({:}\lambda{:} b)\,
    \sum_{k=0}^{\infty}\frac{(a\,\lambda)^{k}}{k!}Z^{:k:}(x).
    \label{eq:ExpMapChgSeries}
\end{equation}

\subsection{Immersion}\label{sec:Immers}

By inspecting the map
\begin{eqnarray}
    \mathbb{C} & \to & \mathbb{C}
    \nonumber  \\
    z & \mapsto & \frac{1+z}{1-z}
    \label{eq:MoebiusMap}
\end{eqnarray}
one sees that the quadrilateral $(1,\frac{1+ z'}{1- z'}, \frac{1+ z}{1-
  z}\frac{1+ z'}{1- z'}, \frac{1+ z}{1- z})$ with $\abs{z}=\abs{z'}<1$ is
mapped to a quadrilateral whose diagonals cross at a right angle. It shows
that the exponential $\Exp(({:}\lambda{:})$ with
$\abs{\lambda}<\frac{2}{\delta}$ maps each rhombus to a convex quadrilateral.
Therefore it is a locally injective map.

\section*{Acknowledgement} \label{sec:Acknowledgements}
This research is supported by the Deutsche Forschungsgemeinschaft in the
framework of Sonderforschungsbereich 288, ``Differential Geometry and Quantum
Physics''. I thank Richard Kenyon for his comments on the draft of this article.

\bibliographystyle{unsrt}
 \bibliography{Exp}

\end{document}